\documentclass[aps,prb,twocolumn,superscriptaddress,showpacs,amsmath,amssymb,longbibliography]{revtex4-2}

\usepackage{amsfonts}
\usepackage{amssymb}
\usepackage{graphicx}
\usepackage{color}

\usepackage{bm}
\usepackage{braket}

\usepackage{hyperref}

\begin{document}

\title{Zipper Entanglement Renormalization for Free Fermions}

\author{Sing Lam Wong}

\author{Ka Chun Pang}

\author{Hoi Chun Po}
\email{hcpo@ust.hk}
\affiliation{Department of Physics, Hong Kong University of Science and Technology, Clear Water Bay, Hong Kong, China}

\date{\today}
\begin{abstract}
Entanglement renormalization refers to a sequence of real-space coarse-graining transformations in which short-range entanglement on successively longer length scales are distilled out. In this work, we introduce a state-based approach, ``zipper entanglement renormalization'' (ZER), for free-fermion systems. 
The name derives from a unitary we construct at every renormalization step, dubbed the zipper, which unzips the state into an approximate tensor product between a short-range entangled state and a renormalized one carrying the longer-range entanglement.
By successively performing ZER on the renormalized states, we obtain a unitary transformation of the input state into a state that is approximately factorized over the emergent renormalization spacetime.
As a demonstration, we apply ZER to one-dimensional models and show that it efficiently disentangles the ground states of the Su-Schrieffer-Heeger model, a scale-invariant critical state, as well as a more general gapless state with two sets of Fermi points.
\end{abstract}

\maketitle

\section{\label{sec:level1} Introduction}
Specifying a generic quantum many-body state is computationally hard due to the exponential vastness of the many-body Hilbert space. Yet, ground states of local Hamiltonians are far from being generic and could admit efficient descriptions which facilitate the computation of physical observables. Tensor network (TN) states, exemplified in one spatial dimension (1D) by the matrix product states (MPS) \cite{MPS_Verstraete_2004,MPS_Cirac_2006,MPS_Verstraete_2006,MPS_Verstraete_2008,MPS_Cirac_2009,MPS_Schuch_2011,MPS_Cirac_2021} and the multiscale entanglement renormalization anstaz (MERA)\cite{MERA_Vidal_2007,MERA_Vidal_2008,MERA_Vidal2009,MERA_Evenbly2009, MERA_Evenbly2010_2,MERA_Evenbly_2010_3,MERA_Evenbly2010,MERA_Silvi_2010,MERA_Evenbly2011,MERA_Evenbly2011,MERA_Evenbly_2014,MERA_Evenbly2015}, follow such philosophy and relate the entanglement structure of the physical states to the architectures of the TN.

In contrast, free-fermion states, which arise naturally as the ground states of bilinear fermionic Hamiltonians, are amenable to classical simulation due to their integrability. They can, nevertheless, be highly entangled, as is the case of a metal with gapless excitation around the Fermi surfaces. Progress on understanding the TN description of free-fermion states could therefore lend insights into the development of TN architectures for highly entangled states, and the TN representations could also serve as natural starting points for the study of strongly correlated electrons \cite{GMPS_schuch_2019,GMPS_Tu_2020,GMPS_Niu_2021, GMPS_Tu_2022, GMPS_Cirac_2022}.

Existing methods for more general interacting systems, like MPS and MERA, have already been applied to the study of free-fermion states \cite{FG_2015,GMPS_schuch_2019,GMPS_Tu_2020,GMPS_Niu_2021, GMPS_Tu_2022, GMPS_Cirac_2022}. In this work, we follow an alternative ``state-based'' paradigm and propose a scheme, dubbed ``zipper entanglement renormalization'' (ZER), which decomposes a given free-fermion state into smaller, local building blocks akin to a TN. 
As suggested by its name, ZER is based on entanglement renormalization \cite{MERA_Vidal_2007,MERA_Vidal_2008} and, similar to MERA, entanglement on successively longer length scales is distilled out in the form of short-range entangled (SRE) states as the renormalization-group (RG) time increases. 

\begin{figure*}[thp!]
\centering
\includegraphics[scale=0.2]{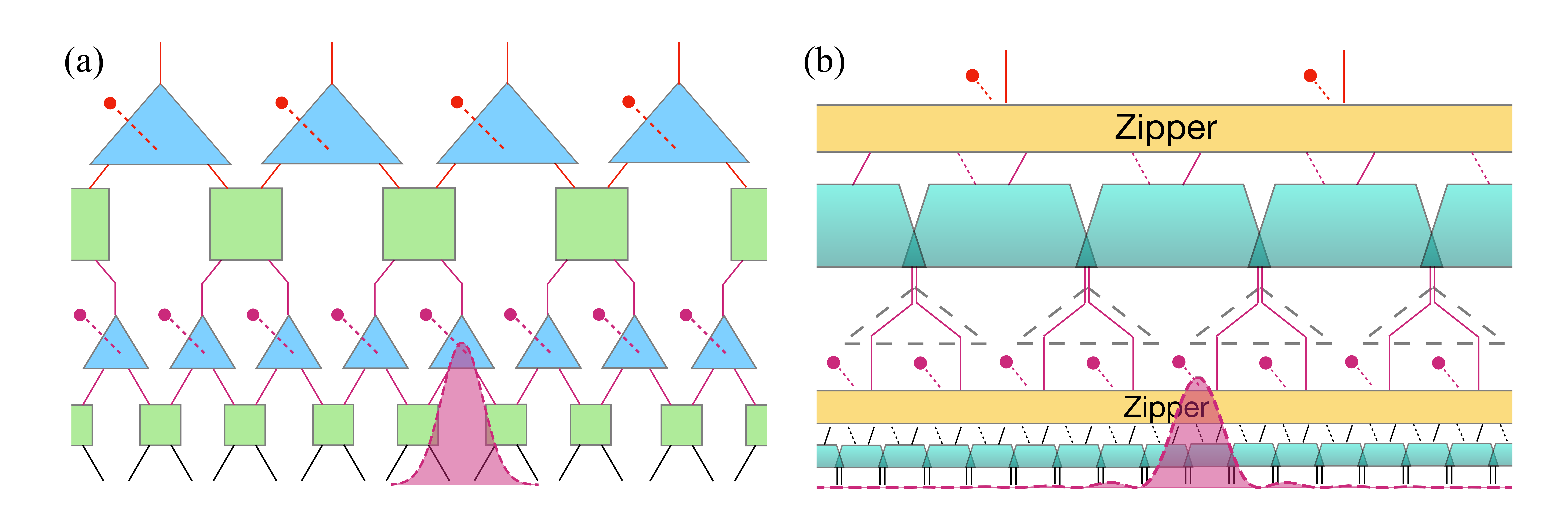}
\caption{Comparison between MERA and ZER.
(a) In MERA, short-range entanglement in the current RG level is distilled out through the successive use of disentagnlers (green squares) and isometries (blue triangles). From the circuit structure, we see that a locally distilled state (terminated dashed leg) corresponds to a state supported on four sites (shaded curve). (b) In ZER, we combine the results of the local distillation (cyan trapezoids) to identify globally defined short-range entangled states which could be recognized as an atomic insulator formed by the filling and emptying of exponentially localized Wannier functions (shaded curve). The transformation to this quasi-local basis is achieved through a unitary we dub the ``zipper'' (yellow block). Longer-range entanglement is carried by the modes associated to the solid legs, and the renormalization step is completed after we block the modes into a super-lattice with an expanded length scale (dashed triangles).
\label{fig:MERAvZER}}  
\end{figure*}

In ZER, the identification of short-range entanglement is achieved through a Hermitian operator, which we call the ``distiller Hamiltonian,'' that spectrally separates short- and long-range entanglement. The Wannier functions of the distiller Hamiltonian can be separated into two sets, which we call ``frozen'' and ``courier.'' 
We refer to the unitary operator performing the basis rotation as the ``zipper,'' in the sense that it approximately unzips the input state into the product of a SRE state defined on the frozen modes, and a renormalized state defined on the courier modes. In other words, the zipper separates the short- and long-range entanglement present in the input state at the current RG scale. Running the RG flow in reverse, the zipper zips up a given state with a product state to produce a similar state defined on a bigger system with, for instance, a doubled system size \cite{RG_Swingle&McGreevy,RG_gapless_Swingle&McGreevy,RG_Swingle_2016}

The SRE states distilled out using ZER are generally not deformable into a strictly local product state using a finite-depth quantum circuit. Instead, in our femrionic context the SRE states distilled out could be interpreted as an atomic insulator formed by the filling of quasi-local Wannier functions, which are not compact and have exponentially decaying tails. This differentiates ZER from other recent discussions on the circuit-based entanglement renormalization of free-fermion states \cite{FG_2015}, like wavelet-based approaches \cite{Wavelet_Evenbly_2016,Wavelet_Evenbly_2018,Wavelet_Haegeman_2018}.

While we envision the ZER scheme to work in any spatial dimension, in this work we focus on 1D free-fermion states as a proof of principle. For a typical 1D critical state defined on a system of length $L$, we show that ZER successfully distills out SRE states and the RG terminates in $\log L$ time, as in other existing schemes like MERA. In contrast, for the ground state of the Su-Schrieffer-Heeger (SSH) model the ZER terminates in $\mathcal O(1)$ time, consistent with the existence of an efficient MPS description for the state. The versatility of the ZER scheme also allows us to treat more realistic models beyond the scale-invariant critical state typically considered \cite{FG_2015,Wavelet_Evenbly_2016, Wavelet_Haegeman_2018}, as we demonstrate with a gapless state away from half filling and featuring two sets of Fermi points.

\section{Zipper entanglement renormalization}
Let $|\Psi \rangle$ be the ground state of a 1D free-fermion Hamiltonian $\hat H$ defined on a finite ring with sites $x=0,\dots, L-1$ for an even $L$. 
For simplicity, we subject the system to a periodic boundary condition and assume translation invariance $x \mapsto x+1 \mod L$.
Let $\hat c_{x,\alpha}$ be the fermion annihilation operators for an orbital labeled by $\alpha$ at site $x$, and we define the correlation matrix $C_{x\alpha, y \beta} \equiv \langle \Psi | \hat c_{x,\alpha}^\dagger \hat c_{y,\beta} | \Psi \rangle$. The correlation matrix determines the reduced density matrix over any subset of sites \cite{Free_fermion_Peschel_2003}, and as such it encodes the entanglement structure of $|\Psi \rangle$ \cite{Free_fermion_Lieb_1961,free_fermion_Peschel_1999,Free_fermion_Peschel_2001,Free_fermion_Cheng_2002,Free_fermion_Peschel_2003,Free_fermion_Vidal_2003,free_fermion_Peschel_2004,Free_fermion_Cheong_2004,free_fermion_Latorre_2009,Free_fermion_Peschel_2009,Free_fermion_Fidkowski_2010}. Indeed, as is the case in existing entanglement renormalization schemes, ZER seeks to systematically inspect correlation matrices defined on successively larger subregions of the system (together with suitable basis rotations) and discover SRE degrees of freedom which could be distilled out at each RG scale. 

Before discussing the details of the ZER scheme, we provide an overview on the procedure. Each RG operation in ZER is composed of three steps (Fig.\ \ref{fig:MERAvZER}): first, we inspect a local region to identify fermionic modes which only contribute entanglement at the current RG scale; second, we combine the results from the local inspection to identify globally defined SRE states which should be distilled out; lastly, we identify the modes carrying longer-range entanglement, and block them into a superlattice with an enlarged lattice constant. The entanglement renormalization is then repeated and it terminates when the system size is reduced to a small $\mathcal O(1)$ number or when the state has been renormalized to a SRE state at an RG step.

\subsection{Local distiller}
The first step in ZER is to identify the short-range entanglement in the input state at the present RG scale. This is achieved by considering the reduced density matrix in a subregion $R$, say consisting of only the first two sites $x=0,1$. The reduced density matrix $\hat \rho^R = e^{- \hat H_{\rm ent}^R}/ {\rm Tr}\big( \exp (- \hat H_{\rm ent}^R) \big)$ is determined by the entanglement Hamiltonian, $\hat H_{\rm ent}^R$, which is also a fermion biinear \cite{Free_fermion_Cheng_2002,Free_fermion_Cheong_2004,Free_fermion_Fidkowski_2010,free_fermion_Latorre_2009,Free_fermion_Lieb_1961,free_fermion_Peschel_1999,Free_fermion_Peschel_2001,Free_fermion_Peschel_2003,free_fermion_Peschel_2004,Free_fermion_Peschel_2009,Free_fermion_Vidal_2003}. The single-particle entanglement Hamiltonian $h^{R}_{ent}$ is related to the restriction of the correlation matrix $R$ through $\left( h^{R}_{ent} \right)^T =   \ln \left(  (1-C^{R})/C^{R}  \right)$, and so its eigenstates and eigenvalues are fully determined by that of $C^R$. Denoting the eigenvalues of $C^R$ by $\xi_i$, which are bounded by $0\leq \xi_i \leq 1$, the von Neumann entanglement entropy $S^R$ of $|\Psi \rangle$ with respect to the bipartition of $R$ and its complement is given by $S^R = -\left (\sum_{i} \xi_{i} \ln \xi_{i}  + (1 - \xi_{i}) \ln(1 - \xi_{i}) \right)$ \cite{Free_fermion_Cheng_2002,Free_fermion_Cheong_2004,Free_fermion_Fidkowski_2010,free_fermion_Latorre_2009,Free_fermion_Lieb_1961,free_fermion_Peschel_1999,Free_fermion_Peschel_2001,Free_fermion_Peschel_2003,free_fermion_Peschel_2004,Free_fermion_Peschel_2009,Free_fermion_Vidal_2003}.

From the expression for $S^R$, we see that eigenmodes of $C^R$ with eigenvalues $\xi_i\simeq 0$ or $1$ have negligible contribution.
We refer to these modes as the ``frozen modes,'' which represent nearly disentangled degrees of freedom \cite{FG_2015}. 
To see why, we notice that the eigenvalues of $C^R$ can be recongized as the fermion occupation numbers of the associated eigenmode, and therefore these modes have little quantum fluctuation as they are essentially frozen to nearly empty or filled.
More precisely, let $\epsilon>0$ be a small numerical threshold. We say an eigenvector $|\phi \rangle_i$ of $C^R$ corresponds to a filled mode if its associated eigenvalue satisfies $\xi_i > 1 - \epsilon$. Similarly, we say $|\phi \rangle_i$ is an empty mode if $\xi_i < \epsilon$. We can separate these frozen modes from the rest through a local, single-particle Hamiltonian
\begin{equation}
    {h}_{\text{distill}}^R = \sum_{i} \ket{\phi_{e}}_{i} \bra{\phi_{e}}_{i} -\sum_{j} \ket{\phi_{f}}_{j} \bra{\phi_{f}}_{j},
\end{equation}
where $\ket{\phi_{f/e}}$ corresponds to the local eigenstates of filled and empty modes in $R$. 
Note that, by design, ${h}_{\text{distill}}^R$ vanishes in the subspace spanned by the non-frozen modes.

\subsection{Global distiller}
The preceding discussion seeks to identify fermion modes localized in $R$ which are frozen in $|\Psi\rangle$. Such modes correspond to correlations in $|\Psi \rangle$ which are confined to the size of $R$, and as such could be viewed as mediating short-range entanglement. By translation invariance, the equivalent frozen modes in other parts of the system should also be distilled out. If we consider non-overlapping regions like $R_{2x} = [2x,2x+1]$ for $x = 0,1,\dots, L/2$, we can distill out all of their respective frozen modes into a product state. Yet, alternative regions like $R_1 = [1,2]$ are also equivalent to $R_0$, and so frozen modes therein should also be treated in the present RG step. Indeed, if we only distill out frozen modes from the non-overlapping subregions $\{ R_{2x}\}$, it will be equivalent to performing a MERA on $|\Psi \rangle$ with only isometries but no disentanglers. This is problematic since, with such a circuit, some of the shortest-scale entanglement would have to propagate to late RG times and this demands an unbounded bond dimension in the thermodynamic limit for a critical state.

\begin{figure}[h]
\centering
\includegraphics[width=0.3 \textwidth]{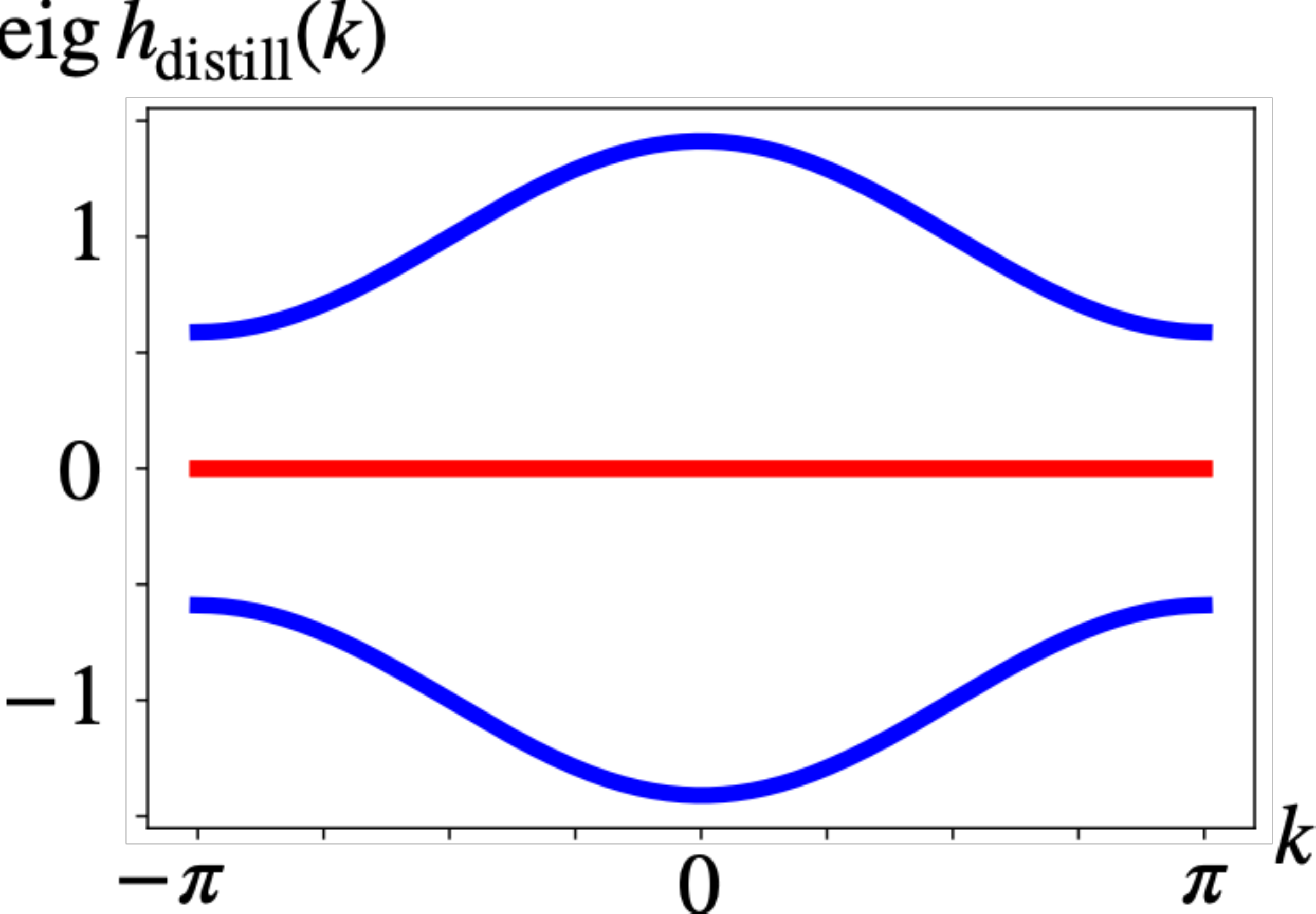}
\caption{An example of the spectral separation between the degrees of freedom mediating short-range entanglement (frozen modes in blue) from the ones carrying longer-range entanglement (courier modes in red) using the global distiller Hamiltonian.}
\label{fig:h_distill}
\end{figure}

The handling of short-range entanglement in the subregions $\{ R_{2x+1}\}$ is a key distinction between ZER and existing approaches like MERA. 
In applying MERA to $|\Psi \rangle$, the entanglement well-localized in the subregions $\{ R_{2x+1}\}$ is handled by the disentanglers, which are then distilled out together with that in $\{ R_{2x}\}$ through the isometries. In contrast, in ZER we treat all the local subregions $\{R_{x} : x=0,1,\dots, L-1\}$ on equal footing. This is achieved by simply summing up the local distiller Hamiltonian into a global one: $h_{\rm distill} = \sum_{x=0}^{L-1} h_{\rm distill}^{R_{x}}$.
Importantly, subregions like $R_0 = [0,1]$ and $R_1 = [1,2]$ overlap, and so the eigenvalues of $h_{\rm distill}$ cease to be $\pm 1$ and $0$. Nevertheless, $h_{\rm distill}$ serves to spectrally separate the frozen modes from those carrying longer-range entanglement, which we dub the ``courier'' modes. 
Furthermore, with the assumed translation invariane, $h_{\rm distill}$ can be diagonalized in the momentum space similar to a regular tight-binding Hamiltonian.
An example is shown in Fig.\ \ref{fig:h_distill}, in which we see that the spectrum of $h_{\rm distill}$ is composed of three sets of bands. The states corresponding to the bands with positive eigenvalues could be reconciled with the empty modes, whereas the ones with negative eigenvalues correspond to the filled modes. These frozen modes are separated from the courier modes, which form the null space of $h_{\rm distill}$ and therefore have exactly zero eigenvalues.

\subsection{Zipper Unitary Transformation}
In line with the general philosophy of entanglement renormalization, we now seek to distill out short-range entanglement from the input state $|\Psi\rangle$. 
More precisely, we claim there exists a quasi-local unitary $\hat U_{\rm zipper}$ such that
\begin{equation}\begin{split}\label{eq:U_zipper}
\hat U_{\text{zipper}} | \Psi \rangle &\approx | \Psi_{\text{courier}} \rangle \otimes | \Psi_{\text{filled}} \rangle \otimes | \Psi_{\text{empty}} \rangle,
\end{split}\end{equation}
where the approximate equality refers to a near factorization of $ \hat U_{\text{zipper}}| \Psi \rangle$ between the courier, filled, and empty modes. 
By quasi-local, we mean that $\hat U_{\rm zipper}$ transformed a local operator to a local one up to an exponentially decaying tail.
Here, $| \Psi_{\text{filled}} \rangle$ and $| \Psi_{\text{empty}} \rangle$ are real-space product states and $| \Psi_{\text{courier}} \rangle$ can be interpreted as the renormalized version of $|\Psi \rangle$ after the frozen modes are distilled out.
We note, however, that the presence of each of the three type of modes is ultimately controlled by the property of the local distillers. For instance, if no state satisfies the local distillation threshold at the present RG step, then $h_\text{distill}=0$ and all modes are labeled as courier, i.e., $\hat U_\text{zipper} = \hat 1$ is trivial. Alternatively, it could also be that all modes are distilled out and no courier mode is left, in which case $|\Psi_\text{courier}\rangle =1$. It is also possible that there are only filled modes but no empty modes, or vice versa.

The extent to which such a factorization is a good approximation can be quantified by the entanglement entropy with respect to the corresponding entanglement cuts. In the local distillation step, the entanglement entropy is controlled by the threshold $
\epsilon$ used in defining the frozen modes, and so locally the entanglement entropy involved in distilling $z$ frozen modes is bounded by $z S(\epsilon)$, where $S(\epsilon) = - \epsilon \log \epsilon$. If we consider distilling from $L$ non-overlapping regions, the corresponding bound becomes $S_{\text{frozen} | \text{courier}} \leq z L S(\epsilon)$, i.e., while it is a volume law, as expected from the extended nature of the ``cut,'' its coefficient is controlled by the small threshold $\epsilon$. For instance, with $\epsilon = 10^{-4}$, the entanglement entropy is suppressed by $S/(zL) \leq S(\epsilon) \simeq 10^{-3}$. Yet, in ZER we distill frozen modes from overlapping regions, and as such the local frozen modes are not orthogonal to each other. Nevertheless, as shown in Appendix \ref{app:bound}, a similar entanglement bound still holds, but now the coefficient is further controlled by the size of the spectral gap of $h_\text{distill}$. If the gap closes, it indicates a failure of separating the frozen modes from the courier modes. Should that happens, we also reject the RG step and treat all modes as courier.

While the entanglement entropy, and hence the near factorization in Eq.\ \eqref{eq:U_zipper}, is independent of the choice of basis within each set of modes, it is important to make sure $\hat U_\text{zipper}$ also maintains a sense of locality; otherwise we will lose the ability to perform further real-space entanglement RG steps on the renormalized state. In ZER, this locality requirement is fulfilled through the use of Wannier functions from band theory \cite{Wannier_Wannier1937,Wanier_PhysRevB.26.4269,Wannier_PhysRevB.56.12847,Wannier_RevModPhys.84.1419,Wannier_Bradlyn_2022}. 
Wannierization (Appendix \ref{app:WF}) can be viewed as the inverse Fourier transformation of a set of delocalized Bloch wave functions, corresponding to an isolated set of energy bands, back into exponentially localized real-space wave functions. 
The basis rotation from the original physical basis to the Wannier basis is specified by a single-particle unitary
\begin{equation}
u_{\text{zipper}} = 
\left(
\begin{array}{ccc}
\psi_{\text{filled}} | & \psi_{\text{courier}} | & \psi_{\text{empty}}
\end{array}
\right),
\end{equation}
where $\psi_{\text{filled}}$,~$\psi_{\text{courier}}$, and $\psi_{\text{empty}}$ are matrices with columns corresponding to the Wannier functions of the filled, courier and empty modes respectively.
As a consequence of the near factorization, the rotated correlation matrix takes the form
\begin{equation} \label{eq:C_trans}
C'^T = u^{\dagger}_{\text{zipper}} C^T u_{\text{zipper}} \approx  
\left(
\begin{array}{ccc} 
\mathbf{\mathbf{1}} & & \\ 
 & C_{\text{courier}}^T  & \\ 
  &  & \mathbf{0}
\end{array}
\right).
\end{equation}
As such, we see that, in the zipper basis, $|\Psi\rangle$ is a near product state between the fully filled state among the filled modes, the vacuum in the empty modes, and a renormalized state with correlation matrix $C_\text{courier}$ in the courier modes.
Furthermore, due to the quasi-local nature of the Wannier functions, the SRE state we distilled out is also quasi-local. 

To continue with the entanglement renormalization, we approximate $\hat U_{\rm zipper} |\Psi\rangle $ by taking the leading Schmidt state across each of the entanglement cuts involved. In our free-fermion context, this is equivalent to replacing each of the sub-blocks on the right-hand side of Eq.\ \eqref{eq:C_trans} by a projector, i.e., we first diagonalize each of the sub-blocks and, due to the near factorization, each of the eigenvalues will be close to $0$ or $1$. To arrive at a projector, we deform these eigenvalues to $0$ and $1$ correspondingly.
Lastly, we block the courier modes in, for instance, two unit cells into one new super site such that the lattice constant is doubled. This defines a new state $|\Psi'\rangle$ living on the Hilbert space of the courier modes, which serves as the starting point of the next ZER step. By composing the zipper unitaries across all RG levels, we have $\hat U_{\text{ZER}} =  \hat U_{\text{zipper}}^{[k]} \cdots \hat U_{\text{zipper}}^{[2]} \hat U_{\text{zipper}}^{[1]} $ and, correspondingly, we obtain the approximation
\begin{equation}
\begin{split}\label{eq:ZER_approx}
    \hat U_{\text{ZER}} | \Psi \rangle 
    &\approx | \Psi_{\text{core}} \rangle \otimes (|\Psi_e^{[k]}\rangle \otimes \cdots \otimes |\Psi_e^{[1]}\rangle )\\ 
    &~~~~~~~~~~~~~~~~~~ \otimes (|\Psi_f^{[k]}\rangle \otimes \cdots \otimes |\Psi_f^{[1]}\rangle  ),
\end{split}
\end{equation}
where $|\Psi_f^{[i]}\rangle$ and $|\Psi_e^{[i]}\rangle$ are the SRE states distilled out in the $i$-th RG step, and $| \Psi_{\text{core}} \rangle$ is the state found at the end of the RG process. As with a usual RG scheme, $| \Psi_{\text{core}} \rangle$  encodes the long-distance properties of the state.

\begin{figure*}[thp]
  {\includegraphics[width=1 \textwidth]{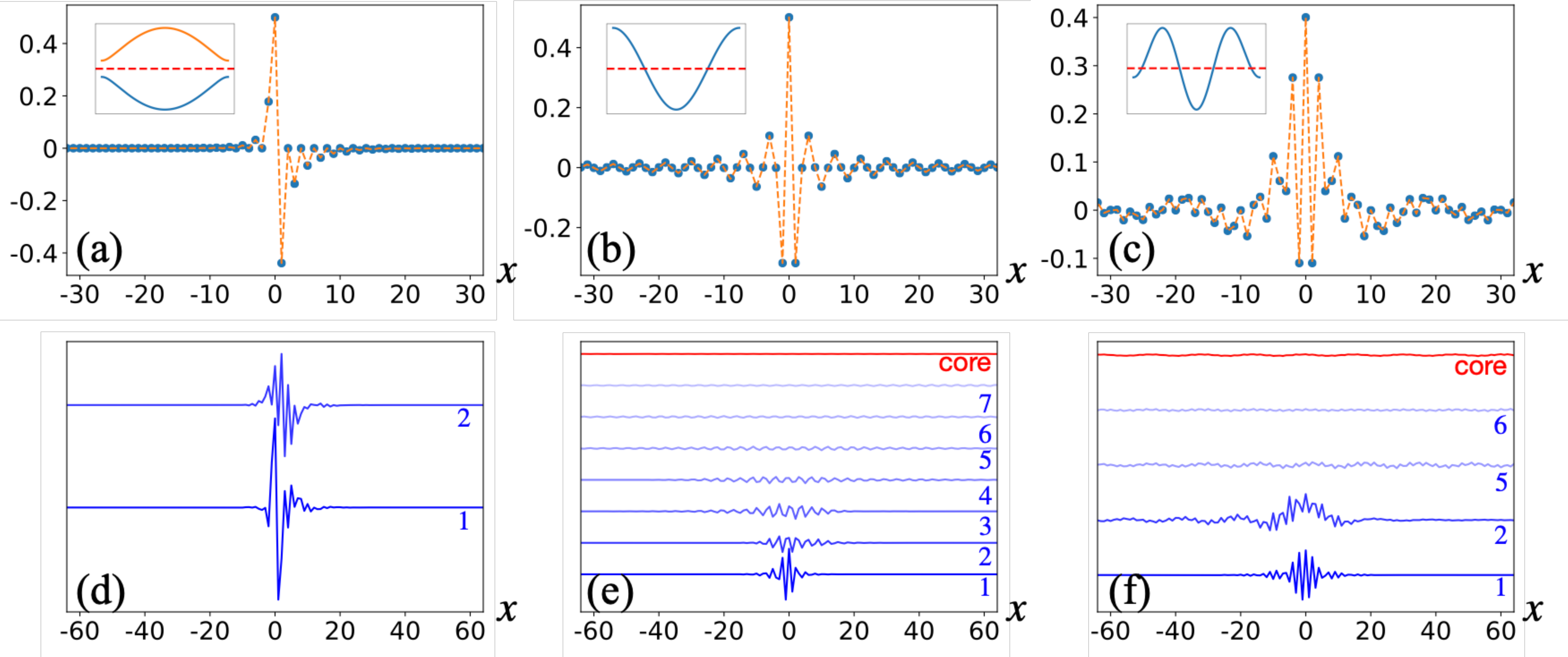}} 
  \caption{Two-point correlations $\langle \hat c_x^\dagger \hat c_0 \rangle$. (a-c) Comparison of the ZER approximation with the input taken as the ground states of three models: (a) the Su-Schrieffer–Heeger Model, (b) the nearest-neighbor half-filled metal, and (c) an extended model at a filling of $0.4$ per site and two electron pockets. The inset shows the energy band (solid line) and the Fermi level (dashed line) defining the input ground states. (d-f) Decomposition of the contributions from different RG levels in ZER. Note that, in (f) the traces for the third and fourth RG steps are missing as only empty modes are distilled out at those steps. 
  \label{fig:corr}
  }
\end{figure*}

\begin{figure}[thp]
  \includegraphics[width=\linewidth]{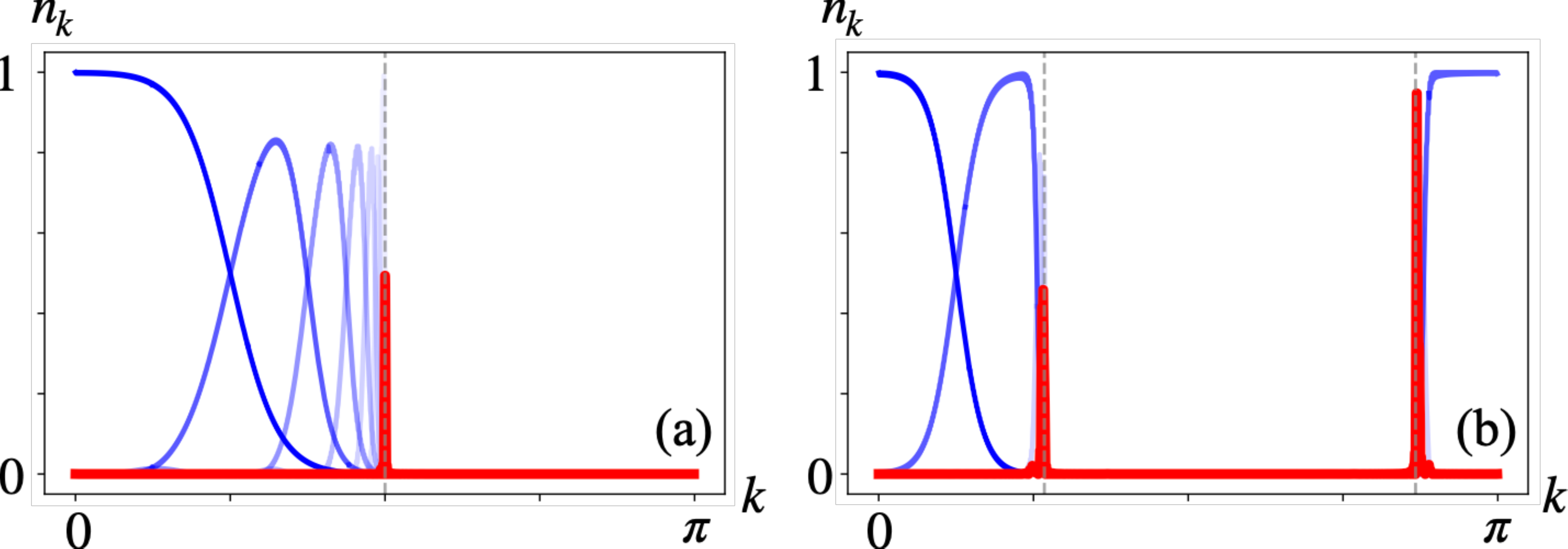}
\caption{Momentum occupation $n_k$ across RG levels. The results for $-k$ (not shown) are the same as those for $+k$.
(a) Nearest-neighbor model. (b) Extended model. Blue traces correspond to filled modes and lighter traces are results from later RG time.
In both cases, ZER systematically distills out the high-energy degrees of freedom and the core modes (red) reside at the Fermi point(s) indicated by dashed line. 
\label{fig:k_occ}} 
\end{figure}

\section{Examples} 
As a proof of principle, we now apply the ZER scheme to perform entanglement renormalization on free-fermion states with varying degrees of entanglement. We consider the ground states of three models of spinless electrons in 1D: (i) the SSH model; (ii) a one-band model with only nearest-neighbor hopping at half-filling; and (iii) an extended one-band model with second nearest-neighbor hopping and at a filling of $0.4$ electron per site. To benchmark the performance of ZER, in Fig.\ \ref{fig:corr} we compare the correlation function $\langle \hat c_i^\dagger \hat c_j \rangle$ obtained from the ZER approximation with the original one computed using the actual ground states of the Hamiltonians. As one can see, there is no visible differences in all cases, which justifies the approximate equality in Eq.\ \ref{eq:ZER_approx}. Furthermore, when we decompose the ZER results into the different RG levels, we see that, as expected, early RG times correspond to short-range correlations, and the long-distance oscillations characteristic of a metal is carried by the core modes. Notice that, for the SSH model, the ZER terminates quickly after two steps, as one expects from the SRE nature of the state.

We emphasize that ZER can handle both the typical critical state in the nearest-neighbor model as well as more general gapless states, like the ground state of the extended model we considered. Further insight could be gleaned by considering the momentum occupation $\langle \Psi_e^{[i]}| \hat c_k^\dagger \hat c_k |\Psi_e^{[i]}\rangle$ at different RG scales (Fig.\ \ref{fig:k_occ}). For the nearest-neighbor model at half-filling, as ZER runs the distilled modes correspond to modes at successively higher momentum until the system reaches the Fermi momentum at $\pi/2$. This coincides readily with the physical picture that the modes distilled out at early RG time are high-energy states far away from the Fermi points. In contrast, for the extended model we see that the frozen modes distilled out at any RG step are generally mixed between the two distinct electron pockets. Nevertheless, the RG progresses towards the two sets of Fermi momenta and the core states are again low-energy modes at the Fermi points.

\section{Conclusion}
In this work, we introduce ZER as a scheme for decomposing a free-fermion state into short-range entangled states defined on successively longer length scales. Correspondingly, the long-distance properties of the state, like the power-law decaying tail of the correlation function $\langle \hat c^\dagger_i \hat c_j \rangle$, are encoded in a small number of ``core'' degrees of freedom. Beyond reproducing the ability to capture a scale invariant critical state \cite{MERA_Evenbly2009,MERA_Evenbly2010_2,MERA_Evenbly2010,MERA_Evenbly2011,MERA_Evenbly2015,MERA_Vidal_2007,MERA_Vidal_2008,MERA_Vidal2009,Wavelet_Evenbly_2016,Wavelet_Evenbly_2018,Wavelet_Haegeman_2018}, we demonstrate the flexibility of ZER with an extended model away from half-filling and featuring two electron pockets. In contrast to MERA \cite{MERA_Evenbly2009, MERA_Evenbly2010_2,MERA_Evenbly_2010_3,MERA_Evenbly2010,MERA_Evenbly2011,MERA_Evenbly2011,MERA_Evenbly2015,MERA_Vidal_2007,MERA_Vidal_2008,MERA_Vidal2009,MERA_Evenbly_2014,MERA_Silvi_2010} or the more recently introduced wavelet-based approaches \cite{Wavelet_Evenbly_2016,Wavelet_Evenbly_2018,Wavelet_Haegeman_2018}, the quasi-local nature of the short-range entangled states distilled out with ZER suggests it might not admit an exact quantum circuit description, although a more general tensor-network operator description for the zipper unitary is anticipated \cite{MP_rep}.
The ZER approach is expected to apply to a much more relaxed setup, including, for instance, systems in higher spatial dimension, and it might serve as a starting point for the construction of correlated fermionic tensor network states \cite{GMPS_schuch_2019,GMPS_Tu_2020,GMPS_Niu_2021, GMPS_Tu_2022, GMPS_Cirac_2022}.

\begin{acknowledgements}
This work is supported by the RGC and the Croucher Foundation through 26308021 and CF21SC01.
We appreciate fruitful discussions with Kangle Li and Yanbai Zhang.  
\end{acknowledgements}

\bibliography{references} 

\providecommand{\noopsort}[1]{}\providecommand{\singleletter}[1]{#1}%
\begin{thebibliography}{47}%
\makeatletter
\providecommand \@ifxundefined [1]{%
 \@ifx{#1\undefined}
}%
\providecommand \@ifnum [1]{%
 \ifnum #1\expandafter \@firstoftwo
 \else \expandafter \@secondoftwo
 \fi
}%
\providecommand \@ifx [1]{%
 \ifx #1\expandafter \@firstoftwo
 \else \expandafter \@secondoftwo
 \fi
}%
\providecommand \natexlab [1]{#1}%
\providecommand \enquote  [1]{``#1''}%
\providecommand \bibnamefont  [1]{#1}%
\providecommand \bibfnamefont [1]{#1}%
\providecommand \citenamefont [1]{#1}%
\providecommand \href@noop [0]{\@secondoftwo}%
\providecommand \href [0]{\begingroup \@sanitize@url \@href}%
\providecommand \@href[1]{\@@startlink{#1}\@@href}%
\providecommand \@@href[1]{\endgroup#1\@@endlink}%
\providecommand \@sanitize@url [0]{\catcode `\\12\catcode `\$12\catcode
  `\&12\catcode `\#12\catcode `\^12\catcode `\_12\catcode `\%12\relax}%
\providecommand \@@startlink[1]{}%
\providecommand \@@endlink[0]{}%
\providecommand \url  [0]{\begingroup\@sanitize@url \@url }%
\providecommand \@url [1]{\endgroup\@href {#1}{\urlprefix }}%
\providecommand \urlprefix  [0]{URL }%
\providecommand \Eprint [0]{\href }%
\providecommand \doibase [0]{https://doi.org/}%
\providecommand \selectlanguage [0]{\@gobble}%
\providecommand \bibinfo  [0]{\@secondoftwo}%
\providecommand \bibfield  [0]{\@secondoftwo}%
\providecommand \translation [1]{[#1]}%
\providecommand \BibitemOpen [0]{}%
\providecommand \bibitemStop [0]{}%
\providecommand \bibitemNoStop [0]{.\EOS\space}%
\providecommand \EOS [0]{\spacefactor3000\relax}%
\providecommand \BibitemShut  [1]{\csname bibitem#1\endcsname}%
\let\auto@bib@innerbib\@empty
\bibitem [{\citenamefont {Verstraete}\ \emph {et~al.}(2004)\citenamefont
  {Verstraete}, \citenamefont {Porras},\ and\ \citenamefont
  {Cirac}}]{MPS_Verstraete_2004}%
  \BibitemOpen
  \bibfield  {author} {\bibinfo {author} {\bibfnamefont {F.}~\bibnamefont
  {Verstraete}}, \bibinfo {author} {\bibfnamefont {D.}~\bibnamefont {Porras}},\
  and\ \bibinfo {author} {\bibfnamefont {J.~I.}\ \bibnamefont {Cirac}},\
  }\bibfield  {title} {\bibinfo {title} {Density matrix renormalization group
  and periodic boundary conditions: A quantum information perspective},\ }\href
  {https://doi.org/10.1103/PhysRevLett.93.227205} {\bibfield  {journal}
  {\bibinfo  {journal} {Phys. Rev. Lett.}\ }\textbf {\bibinfo {volume} {93}},\
  \bibinfo {pages} {227205} (\bibinfo {year} {2004})}\BibitemShut {NoStop}%
\bibitem [{\citenamefont {Perez-Garcia}\ \emph {et~al.}(2006)\citenamefont
  {Perez-Garcia}, \citenamefont {Verstraete}, \citenamefont {Wolf},\ and\
  \citenamefont {Cirac}}]{MPS_Cirac_2006}%
  \BibitemOpen
  \bibfield  {author} {\bibinfo {author} {\bibfnamefont {D.}~\bibnamefont
  {Perez-Garcia}}, \bibinfo {author} {\bibfnamefont {F.}~\bibnamefont
  {Verstraete}}, \bibinfo {author} {\bibfnamefont {M.~M.}\ \bibnamefont
  {Wolf}},\ and\ \bibinfo {author} {\bibfnamefont {J.~I.}\ \bibnamefont
  {Cirac}},\ }\bibfield  {title} {\bibinfo {title} {Matrix product state
  representations}\ }\href {https://doi.org/10.48550/ARXIV.QUANT-PH/0608197}
  {10.48550/ARXIV.QUANT-PH/0608197} (\bibinfo {year} {2006})\BibitemShut
  {NoStop}%
\bibitem [{\citenamefont {Verstraete}\ and\ \citenamefont
  {Cirac}(2006)}]{MPS_Verstraete_2006}%
  \BibitemOpen
  \bibfield  {author} {\bibinfo {author} {\bibfnamefont {F.}~\bibnamefont
  {Verstraete}}\ and\ \bibinfo {author} {\bibfnamefont {J.~I.}\ \bibnamefont
  {Cirac}},\ }\bibfield  {title} {\bibinfo {title} {Matrix product states
  represent ground states faithfully},\ }\bibfield  {journal} {\bibinfo
  {journal} {Physical Review B}\ }\textbf {\bibinfo {volume} {73}},\ \href
  {https://doi.org/10.1103/physrevb.73.094423} {10.1103/physrevb.73.094423}
  (\bibinfo {year} {2006})\BibitemShut {NoStop}%
\bibitem [{\citenamefont {Verstraete}\ \emph {et~al.}(2008)\citenamefont
  {Verstraete}, \citenamefont {Murg},\ and\ \citenamefont
  {Cirac}}]{MPS_Verstraete_2008}%
  \BibitemOpen
  \bibfield  {author} {\bibinfo {author} {\bibfnamefont {F.}~\bibnamefont
  {Verstraete}}, \bibinfo {author} {\bibfnamefont {V.}~\bibnamefont {Murg}},\
  and\ \bibinfo {author} {\bibfnamefont {J.}~\bibnamefont {Cirac}},\ }\bibfield
   {title} {\bibinfo {title} {Matrix product states, projected entangled pair
  states, and variational renormalization group methods for quantum spin
  systems},\ }\href {https://doi.org/10.1080/14789940801912366} {\bibfield
  {journal} {\bibinfo  {journal} {Advances in Physics}\ }\textbf {\bibinfo
  {volume} {57}},\ \bibinfo {pages} {143} (\bibinfo {year} {2008})}\BibitemShut
  {NoStop}%
\bibitem [{\citenamefont {Cirac}\ and\ \citenamefont
  {Verstraete}(2009)}]{MPS_Cirac_2009}%
  \BibitemOpen
  \bibfield  {author} {\bibinfo {author} {\bibfnamefont {J.~I.}\ \bibnamefont
  {Cirac}}\ and\ \bibinfo {author} {\bibfnamefont {F.}~\bibnamefont
  {Verstraete}},\ }\bibfield  {title} {\bibinfo {title} {Renormalization and
  tensor product states in spin chains and lattices},\ }\href
  {https://doi.org/10.1088/1751-8113/42/50/504004} {\bibfield  {journal}
  {\bibinfo  {journal} {Journal of Physics A: Mathematical and Theoretical}\
  }\textbf {\bibinfo {volume} {42}},\ \bibinfo {pages} {504004} (\bibinfo
  {year} {2009})}\BibitemShut {NoStop}%
\bibitem [{\citenamefont {Schuch}\ \emph {et~al.}(2011)\citenamefont {Schuch},
  \citenamefont {P{\'{e} }rez-Garc{\'{\i}}a},\ and\ \citenamefont
  {Cirac}}]{MPS_Schuch_2011}%
  \BibitemOpen
  \bibfield  {author} {\bibinfo {author} {\bibfnamefont {N.}~\bibnamefont
  {Schuch}}, \bibinfo {author} {\bibfnamefont {D.}~\bibnamefont {P{\'{e}
  }rez-Garc{\'{\i}}a}},\ and\ \bibinfo {author} {\bibfnamefont
  {I.}~\bibnamefont {Cirac}},\ }\bibfield  {title} {\bibinfo {title}
  {Classifying quantum phases using matrix product states and projected
  entangled pair states},\ }\bibfield  {journal} {\bibinfo  {journal} {Physical
  Review B}\ }\textbf {\bibinfo {volume} {84}},\ \href
  {https://doi.org/10.1103/physrevb.84.165139} {10.1103/physrevb.84.165139}
  (\bibinfo {year} {2011})\BibitemShut {NoStop}%
\bibitem [{\citenamefont {Cirac}\ \emph {et~al.}(2021)\citenamefont {Cirac},
  \citenamefont {P{\'{e} }rez-Garc{\'{\i}}a}, \citenamefont {Schuch},\ and\
  \citenamefont {Verstraete}}]{MPS_Cirac_2021}%
  \BibitemOpen
  \bibfield  {author} {\bibinfo {author} {\bibfnamefont {J.~I.}\ \bibnamefont
  {Cirac}}, \bibinfo {author} {\bibfnamefont {D.}~\bibnamefont {P{\'{e}
  }rez-Garc{\'{\i}}a}}, \bibinfo {author} {\bibfnamefont {N.}~\bibnamefont
  {Schuch}},\ and\ \bibinfo {author} {\bibfnamefont {F.}~\bibnamefont
  {Verstraete}},\ }\bibfield  {title} {\bibinfo {title} {Matrix product states
  and projected entangled pair states: Concepts, symmetries, theorems},\
  }\bibfield  {journal} {\bibinfo  {journal} {Reviews of Modern Physics}\
  }\textbf {\bibinfo {volume} {93}},\ \href
  {https://doi.org/10.1103/revmodphys.93.045003} {10.1103/revmodphys.93.045003}
  (\bibinfo {year} {2021})\BibitemShut {NoStop}%
\bibitem [{\citenamefont {Vidal}(2007)}]{MERA_Vidal_2007}%
  \BibitemOpen
  \bibfield  {author} {\bibinfo {author} {\bibfnamefont {G.}~\bibnamefont
  {Vidal}},\ }\bibfield  {title} {\bibinfo {title} {Entanglement
  renormalization},\ }\bibfield  {journal} {\bibinfo  {journal} {Physical
  Review Letters}\ }\textbf {\bibinfo {volume} {99}},\ \href
  {https://doi.org/10.1103/physrevlett.99.220405}
  {10.1103/physrevlett.99.220405} (\bibinfo {year} {2007})\BibitemShut
  {NoStop}%
\bibitem [{\citenamefont {Vidal}(2008)}]{MERA_Vidal_2008}%
  \BibitemOpen
  \bibfield  {author} {\bibinfo {author} {\bibfnamefont {G.}~\bibnamefont
  {Vidal}},\ }\bibfield  {title} {\bibinfo {title} {Class of quantum many-body
  states that can be efficiently simulated},\ }\bibfield  {journal} {\bibinfo
  {journal} {Physical Review Letters}\ }\textbf {\bibinfo {volume} {101}},\
  \href {https://doi.org/10.1103/physrevlett.101.110501}
  {10.1103/physrevlett.101.110501} (\bibinfo {year} {2008})\BibitemShut
  {NoStop}%
\bibitem [{\citenamefont {Vidal}(2009)}]{MERA_Vidal2009}%
  \BibitemOpen
  \bibfield  {author} {\bibinfo {author} {\bibfnamefont {G.}~\bibnamefont
  {Vidal}},\ }\href {https://doi.org/10.48550/ARXIV.0912.1651} {\bibinfo
  {title} {Entanglement renormalization: an introduction}} (\bibinfo {year}
  {2009})\BibitemShut {NoStop}%
\bibitem [{\citenamefont {Pfeifer}\ \emph {et~al.}(2009)\citenamefont
  {Pfeifer}, \citenamefont {Evenbly},\ and\ \citenamefont
  {Vidal}}]{MERA_Evenbly2009}%
  \BibitemOpen
  \bibfield  {author} {\bibinfo {author} {\bibfnamefont {R.~N.~C.}\
  \bibnamefont {Pfeifer}}, \bibinfo {author} {\bibfnamefont {G.}~\bibnamefont
  {Evenbly}},\ and\ \bibinfo {author} {\bibfnamefont {G.}~\bibnamefont
  {Vidal}},\ }\bibfield  {title} {\bibinfo {title} {Entanglement
  renormalization, scale invariance, and quantum criticality},\ }\href
  {https://doi.org/10.1103/PhysRevA.79.040301} {\bibfield  {journal} {\bibinfo
  {journal} {Phys. Rev. A}\ }\textbf {\bibinfo {volume} {79}},\ \bibinfo
  {pages} {040301} (\bibinfo {year} {2009})}\BibitemShut {NoStop}%
\bibitem [{\citenamefont {Evenbly}\ and\ \citenamefont
  {Vidal}(2010)}]{MERA_Evenbly2010_2}%
  \BibitemOpen
  \bibfield  {author} {\bibinfo {author} {\bibfnamefont {G.}~\bibnamefont
  {Evenbly}}\ and\ \bibinfo {author} {\bibfnamefont {G.}~\bibnamefont
  {Vidal}},\ }\bibfield  {title} {\bibinfo {title} {Entanglement
  renormalization in noninteracting fermionic systems},\ }\href
  {https://doi.org/10.1103/PhysRevB.81.235102} {\bibfield  {journal} {\bibinfo
  {journal} {Phys. Rev. B}\ }\textbf {\bibinfo {volume} {81}},\ \bibinfo
  {pages} {235102} (\bibinfo {year} {2010})}\BibitemShut {NoStop}%
\bibitem [{\citenamefont {Evenbly}\ \emph {et~al.}(2010)\citenamefont
  {Evenbly}, \citenamefont {Pfeifer}, \citenamefont {Pic{\'{o} }},
  \citenamefont {Iblisdir}, \citenamefont {Tagliacozzo}, \citenamefont
  {McCulloch},\ and\ \citenamefont {Vidal}}]{MERA_Evenbly_2010_3}%
  \BibitemOpen
  \bibfield  {author} {\bibinfo {author} {\bibfnamefont {G.}~\bibnamefont
  {Evenbly}}, \bibinfo {author} {\bibfnamefont {R.~N.~C.}\ \bibnamefont
  {Pfeifer}}, \bibinfo {author} {\bibfnamefont {V.}~\bibnamefont {Pic{\'{o}
  }}}, \bibinfo {author} {\bibfnamefont {S.}~\bibnamefont {Iblisdir}}, \bibinfo
  {author} {\bibfnamefont {L.}~\bibnamefont {Tagliacozzo}}, \bibinfo {author}
  {\bibfnamefont {I.~P.}\ \bibnamefont {McCulloch}},\ and\ \bibinfo {author}
  {\bibfnamefont {G.}~\bibnamefont {Vidal}},\ }\bibfield  {title} {\bibinfo
  {title} {Boundary quantum critical phenomena with entanglement
  renormalization},\ }\bibfield  {journal} {\bibinfo  {journal} {Physical
  Review B}\ }\textbf {\bibinfo {volume} {82}},\ \href
  {https://doi.org/10.1103/physrevb.82.161107} {10.1103/physrevb.82.161107}
  (\bibinfo {year} {2010})\BibitemShut {NoStop}%
\bibitem [{\citenamefont {Corboz}\ \emph {et~al.}(2010)\citenamefont {Corboz},
  \citenamefont {Evenbly}, \citenamefont {Verstraete},\ and\ \citenamefont
  {Vidal}}]{MERA_Evenbly2010}%
  \BibitemOpen
  \bibfield  {author} {\bibinfo {author} {\bibfnamefont {P.}~\bibnamefont
  {Corboz}}, \bibinfo {author} {\bibfnamefont {G.}~\bibnamefont {Evenbly}},
  \bibinfo {author} {\bibfnamefont {F.}~\bibnamefont {Verstraete}},\ and\
  \bibinfo {author} {\bibfnamefont {G.}~\bibnamefont {Vidal}},\ }\bibfield
  {title} {\bibinfo {title} {Simulation of interacting fermions with
  entanglement renormalization},\ }\href
  {https://doi.org/10.1103/PhysRevA.81.010303} {\bibfield  {journal} {\bibinfo
  {journal} {Phys. Rev. A}\ }\textbf {\bibinfo {volume} {81}},\ \bibinfo
  {pages} {010303} (\bibinfo {year} {2010})}\BibitemShut {NoStop}%
\bibitem [{\citenamefont {Silvi}\ \emph {et~al.}(2010)\citenamefont {Silvi},
  \citenamefont {Giovannetti}, \citenamefont {Calabrese}, \citenamefont
  {Santoro},\ and\ \citenamefont {Fazio}}]{MERA_Silvi_2010}%
  \BibitemOpen
  \bibfield  {author} {\bibinfo {author} {\bibfnamefont {P.}~\bibnamefont
  {Silvi}}, \bibinfo {author} {\bibfnamefont {V.}~\bibnamefont {Giovannetti}},
  \bibinfo {author} {\bibfnamefont {P.}~\bibnamefont {Calabrese}}, \bibinfo
  {author} {\bibfnamefont {G.~E.}\ \bibnamefont {Santoro}},\ and\ \bibinfo
  {author} {\bibfnamefont {R.}~\bibnamefont {Fazio}},\ }\bibfield  {title}
  {\bibinfo {title} {Entanglement renormalization and boundary critical
  phenomena},\ }\href {https://doi.org/10.1088/1742-5468/2010/03/l03001}
  {\bibfield  {journal} {\bibinfo  {journal} {Journal of Statistical Mechanics:
  Theory and Experiment}\ }\textbf {\bibinfo {volume} {2010}},\ \bibinfo
  {pages} {L03001} (\bibinfo {year} {2010})}\BibitemShut {NoStop}%
\bibitem [{\citenamefont {Evenbly}\ and\ \citenamefont
  {Vidal}(2011)}]{MERA_Evenbly2011}%
  \BibitemOpen
  \bibfield  {author} {\bibinfo {author} {\bibfnamefont {G.}~\bibnamefont
  {Evenbly}}\ and\ \bibinfo {author} {\bibfnamefont {G.}~\bibnamefont
  {Vidal}},\ }\href {https://doi.org/10.48550/ARXIV.1109.5334} {\bibinfo
  {title} {Quantum criticality with the multi-scale entanglement
  renormalization ansatz}} (\bibinfo {year} {2011})\BibitemShut {NoStop}%
\bibitem [{\citenamefont {Evenbly}\ and\ \citenamefont
  {Vidal}(2014)}]{MERA_Evenbly_2014}%
  \BibitemOpen
  \bibfield  {author} {\bibinfo {author} {\bibfnamefont {G.}~\bibnamefont
  {Evenbly}}\ and\ \bibinfo {author} {\bibfnamefont {G.}~\bibnamefont
  {Vidal}},\ }\bibfield  {title} {\bibinfo {title} {Class of highly entangled
  many-body states that can be efficiently simulated},\ }\bibfield  {journal}
  {\bibinfo  {journal} {Physical Review Letters}\ }\textbf {\bibinfo {volume}
  {112}},\ \href {https://doi.org/10.1103/physrevlett.112.240502}
  {10.1103/physrevlett.112.240502} (\bibinfo {year} {2014})\BibitemShut
  {NoStop}%
\bibitem [{\citenamefont {Evenbly}\ and\ \citenamefont
  {Vidal}(2015)}]{MERA_Evenbly2015}%
  \BibitemOpen
  \bibfield  {author} {\bibinfo {author} {\bibfnamefont {G.}~\bibnamefont
  {Evenbly}}\ and\ \bibinfo {author} {\bibfnamefont {G.}~\bibnamefont
  {Vidal}},\ }\bibfield  {title} {\bibinfo {title} {Tensor network
  renormalization yields the multiscale entanglement renormalization ansatz},\
  }\bibfield  {journal} {\bibinfo  {journal} {Physical Review Letters}\
  }\textbf {\bibinfo {volume} {115}},\ \href
  {https://doi.org/10.1103/physrevlett.115.200401}
  {10.1103/physrevlett.115.200401} (\bibinfo {year} {2015})\BibitemShut
  {NoStop}%
\bibitem [{\citenamefont {Schuch}\ and\ \citenamefont
  {Bauer}(2019)}]{GMPS_schuch_2019}%
  \BibitemOpen
  \bibfield  {author} {\bibinfo {author} {\bibfnamefont {N.}~\bibnamefont
  {Schuch}}\ and\ \bibinfo {author} {\bibfnamefont {B.}~\bibnamefont {Bauer}},\
  }\bibfield  {title} {\bibinfo {title} {Matrix product state algorithms for
  gaussian fermionic states},\ }\href
  {https://doi.org/10.1103/PhysRevB.100.245121} {\bibfield  {journal} {\bibinfo
   {journal} {Phys. Rev. B}\ }\textbf {\bibinfo {volume} {100}},\ \bibinfo
  {pages} {245121} (\bibinfo {year} {2019})}\BibitemShut {NoStop}%
\bibitem [{\citenamefont {Wu}\ \emph {et~al.}(2020)\citenamefont {Wu},
  \citenamefont {Wang},\ and\ \citenamefont {Tu}}]{GMPS_Tu_2020}%
  \BibitemOpen
  \bibfield  {author} {\bibinfo {author} {\bibfnamefont {Y.-H.}\ \bibnamefont
  {Wu}}, \bibinfo {author} {\bibfnamefont {L.}~\bibnamefont {Wang}},\ and\
  \bibinfo {author} {\bibfnamefont {H.-H.}\ \bibnamefont {Tu}},\ }\bibfield
  {title} {\bibinfo {title} {Tensor network representations of parton wave
  functions},\ }\href {https://doi.org/10.1103/PhysRevLett.124.246401}
  {\bibfield  {journal} {\bibinfo  {journal} {Phys. Rev. Lett.}\ }\textbf
  {\bibinfo {volume} {124}},\ \bibinfo {pages} {246401} (\bibinfo {year}
  {2020})}\BibitemShut {NoStop}%
\bibitem [{\citenamefont {Niu}\ \emph {et~al.}(2021)\citenamefont {Niu},
  \citenamefont {Haghshenas}, \citenamefont {Zhang}, \citenamefont {Foss-Feig},
  \citenamefont {Chan},\ and\ \citenamefont {Potter}}]{GMPS_Niu_2021}%
  \BibitemOpen
  \bibfield  {author} {\bibinfo {author} {\bibfnamefont {D.}~\bibnamefont
  {Niu}}, \bibinfo {author} {\bibfnamefont {R.}~\bibnamefont {Haghshenas}},
  \bibinfo {author} {\bibfnamefont {Y.}~\bibnamefont {Zhang}}, \bibinfo
  {author} {\bibfnamefont {M.}~\bibnamefont {Foss-Feig}}, \bibinfo {author}
  {\bibfnamefont {G.~K.-L.}\ \bibnamefont {Chan}},\ and\ \bibinfo {author}
  {\bibfnamefont {A.~C.}\ \bibnamefont {Potter}},\ }\href
  {https://doi.org/10.48550/ARXIV.2112.10810} {\bibinfo {title} {Holographic
  simulation of correlated electrons on a trapped ion quantum processor}}
  (\bibinfo {year} {2021})\BibitemShut {NoStop}%
\bibitem [{\citenamefont {Jin}\ \emph {et~al.}(2022)\citenamefont {Jin},
  \citenamefont {Sun}, \citenamefont {Zhou},\ and\ \citenamefont
  {Tu}}]{GMPS_Tu_2022}%
  \BibitemOpen
  \bibfield  {author} {\bibinfo {author} {\bibfnamefont {H.-K.}\ \bibnamefont
  {Jin}}, \bibinfo {author} {\bibfnamefont {R.-Y.}\ \bibnamefont {Sun}},
  \bibinfo {author} {\bibfnamefont {Y.}~\bibnamefont {Zhou}},\ and\ \bibinfo
  {author} {\bibfnamefont {H.-H.}\ \bibnamefont {Tu}},\ }\bibfield  {title}
  {\bibinfo {title} {Matrix product states for hartree-fock-bogoliubov wave
  functions},\ }\href {https://doi.org/10.1103/PhysRevB.105.L081101} {\bibfield
   {journal} {\bibinfo  {journal} {Phys. Rev. B}\ }\textbf {\bibinfo {volume}
  {105}},\ \bibinfo {pages} {L081101} (\bibinfo {year} {2022})}\BibitemShut
  {NoStop}%
\bibitem [{\citenamefont {Franco-Rubio}\ and\ \citenamefont
  {Cirac}(2022)}]{GMPS_Cirac_2022}%
  \BibitemOpen
  \bibfield  {author} {\bibinfo {author} {\bibfnamefont {A.}~\bibnamefont
  {Franco-Rubio}}\ and\ \bibinfo {author} {\bibfnamefont {J.~I.}\ \bibnamefont
  {Cirac}},\ }\href {https://doi.org/10.48550/ARXIV.2204.02478} {\bibinfo
  {title} {Gaussian matrix product states cannot efficiently describe critical
  systems}} (\bibinfo {year} {2022})\BibitemShut {NoStop}%
\bibitem [{\citenamefont {Fishman}\ and\ \citenamefont
  {White}(2015)}]{FG_2015}%
  \BibitemOpen
  \bibfield  {author} {\bibinfo {author} {\bibfnamefont {M.~T.}\ \bibnamefont
  {Fishman}}\ and\ \bibinfo {author} {\bibfnamefont {S.~R.}\ \bibnamefont
  {White}},\ }\bibfield  {title} {\bibinfo {title} {Compression of correlation
  matrices and an efficient method for forming matrix product states of
  fermionic gaussian states},\ }\bibfield  {journal} {\bibinfo  {journal}
  {Physical Review B}\ }\textbf {\bibinfo {volume} {92}},\ \href
  {https://doi.org/10.1103/physrevb.92.075132} {10.1103/physrevb.92.075132}
  (\bibinfo {year} {2015})\BibitemShut {NoStop}%
\bibitem [{\citenamefont {Swingle}\ and\ \citenamefont
  {McGreevy}(2016{\natexlab{a}})}]{RG_Swingle&McGreevy}%
  \BibitemOpen
  \bibfield  {author} {\bibinfo {author} {\bibfnamefont {B.}~\bibnamefont
  {Swingle}}\ and\ \bibinfo {author} {\bibfnamefont {J.}~\bibnamefont
  {McGreevy}},\ }\bibfield  {title} {\bibinfo {title} {Renormalization group
  constructions of topological quantum liquids and beyond},\ }\href
  {https://doi.org/10.1103/PhysRevB.93.045127} {\bibfield  {journal} {\bibinfo
  {journal} {Phys. Rev. B}\ }\textbf {\bibinfo {volume} {93}},\ \bibinfo
  {pages} {045127} (\bibinfo {year} {2016}{\natexlab{a}})}\BibitemShut
  {NoStop}%
\bibitem [{\citenamefont {Swingle}\ \emph {et~al.}(2016)\citenamefont
  {Swingle}, \citenamefont {McGreevy},\ and\ \citenamefont
  {Xu}}]{RG_gapless_Swingle&McGreevy}%
  \BibitemOpen
  \bibfield  {author} {\bibinfo {author} {\bibfnamefont {B.}~\bibnamefont
  {Swingle}}, \bibinfo {author} {\bibfnamefont {J.}~\bibnamefont {McGreevy}},\
  and\ \bibinfo {author} {\bibfnamefont {S.}~\bibnamefont {Xu}},\ }\bibfield
  {title} {\bibinfo {title} {Renormalization group circuits for gapless
  states},\ }\href {https://doi.org/10.1103/PhysRevB.93.205159} {\bibfield
  {journal} {\bibinfo  {journal} {Phys. Rev. B}\ }\textbf {\bibinfo {volume}
  {93}},\ \bibinfo {pages} {205159} (\bibinfo {year} {2016})}\BibitemShut
  {NoStop}%
\bibitem [{\citenamefont {Swingle}\ and\ \citenamefont
  {McGreevy}(2016{\natexlab{b}})}]{RG_Swingle_2016}%
  \BibitemOpen
  \bibfield  {author} {\bibinfo {author} {\bibfnamefont {B.}~\bibnamefont
  {Swingle}}\ and\ \bibinfo {author} {\bibfnamefont {J.}~\bibnamefont
  {McGreevy}},\ }\bibfield  {title} {\bibinfo {title} {Mixed $s$-sourcery:
  Building many-body states using bubbles of nothing},\ }\href
  {https://doi.org/10.1103/PhysRevB.94.155125} {\bibfield  {journal} {\bibinfo
  {journal} {Phys. Rev. B}\ }\textbf {\bibinfo {volume} {94}},\ \bibinfo
  {pages} {155125} (\bibinfo {year} {2016}{\natexlab{b}})}\BibitemShut
  {NoStop}%
\bibitem [{\citenamefont {Evenbly}\ and\ \citenamefont
  {White}(2016)}]{Wavelet_Evenbly_2016}%
  \BibitemOpen
  \bibfield  {author} {\bibinfo {author} {\bibfnamefont {G.}~\bibnamefont
  {Evenbly}}\ and\ \bibinfo {author} {\bibfnamefont {S.~R.}\ \bibnamefont
  {White}},\ }\bibfield  {title} {\bibinfo {title} {Entanglement
  renormalization and wavelets},\ }\bibfield  {journal} {\bibinfo  {journal}
  {Physical Review Letters}\ }\textbf {\bibinfo {volume} {116}},\ \href
  {https://doi.org/10.1103/physrevlett.116.140403}
  {10.1103/physrevlett.116.140403} (\bibinfo {year} {2016})\BibitemShut
  {NoStop}%
\bibitem [{\citenamefont {Evenbly}\ and\ \citenamefont
  {White}(2018)}]{Wavelet_Evenbly_2018}%
  \BibitemOpen
  \bibfield  {author} {\bibinfo {author} {\bibfnamefont {G.}~\bibnamefont
  {Evenbly}}\ and\ \bibinfo {author} {\bibfnamefont {S.~R.}\ \bibnamefont
  {White}},\ }\bibfield  {title} {\bibinfo {title} {Representation and design
  of wavelets using unitary circuits},\ }\bibfield  {journal} {\bibinfo
  {journal} {Physical Review A}\ }\textbf {\bibinfo {volume} {97}},\ \href
  {https://doi.org/10.1103/physreva.97.052314} {10.1103/physreva.97.052314}
  (\bibinfo {year} {2018})\BibitemShut {NoStop}%
\bibitem [{\citenamefont {Haegeman}\ \emph {et~al.}(2018)\citenamefont
  {Haegeman}, \citenamefont {Swingle}, \citenamefont {Walter}, \citenamefont
  {Cotler}, \citenamefont {Evenbly},\ and\ \citenamefont
  {Scholz}}]{Wavelet_Haegeman_2018}%
  \BibitemOpen
  \bibfield  {author} {\bibinfo {author} {\bibfnamefont {J.}~\bibnamefont
  {Haegeman}}, \bibinfo {author} {\bibfnamefont {B.}~\bibnamefont {Swingle}},
  \bibinfo {author} {\bibfnamefont {M.}~\bibnamefont {Walter}}, \bibinfo
  {author} {\bibfnamefont {J.}~\bibnamefont {Cotler}}, \bibinfo {author}
  {\bibfnamefont {G.}~\bibnamefont {Evenbly}},\ and\ \bibinfo {author}
  {\bibfnamefont {V.~B.}\ \bibnamefont {Scholz}},\ }\bibfield  {title}
  {\bibinfo {title} {Rigorous free-fermion entanglement renormalization from
  wavelet theory},\ }\bibfield  {journal} {\bibinfo  {journal} {Physical Review
  X}\ }\textbf {\bibinfo {volume} {8}},\ \href
  {https://doi.org/10.1103/physrevx.8.011003} {10.1103/physrevx.8.011003}
  (\bibinfo {year} {2018})\BibitemShut {NoStop}%
\bibitem [{\citenamefont {Peschel}(2003)}]{Free_fermion_Peschel_2003}%
  \BibitemOpen
  \bibfield  {author} {\bibinfo {author} {\bibfnamefont {I.}~\bibnamefont
  {Peschel}},\ }\bibfield  {title} {\bibinfo {title} {Calculation of reduced
  density matrices from correlation functions},\ }\href
  {https://doi.org/10.1088/0305-4470/36/14/101} {\bibfield  {journal} {\bibinfo
   {journal} {Journal of Physics A: Mathematical and General}\ }\textbf
  {\bibinfo {volume} {36}},\ \bibinfo {pages} {L205} (\bibinfo {year}
  {2003})}\BibitemShut {NoStop}%
\bibitem [{\citenamefont {Lieb}\ \emph {et~al.}(1961)\citenamefont {Lieb},
  \citenamefont {Schultz},\ and\ \citenamefont
  {Mattis}}]{Free_fermion_Lieb_1961}%
  \BibitemOpen
  \bibfield  {author} {\bibinfo {author} {\bibfnamefont {E.}~\bibnamefont
  {Lieb}}, \bibinfo {author} {\bibfnamefont {T.}~\bibnamefont {Schultz}},\ and\
  \bibinfo {author} {\bibfnamefont {D.}~\bibnamefont {Mattis}},\ }\bibfield
  {title} {\bibinfo {title} {Two soluble models of an antiferromagnetic
  chain},\ }\href
  {https://doi.org/https://doi.org/10.1016/0003-4916(61)90115-4} {\bibfield
  {journal} {\bibinfo  {journal} {Annals of Physics}\ }\textbf {\bibinfo
  {volume} {16}},\ \bibinfo {pages} {407} (\bibinfo {year} {1961})}\BibitemShut
  {NoStop}%
\bibitem [{\citenamefont {Peschel}\ and\ \citenamefont
  {Chung}(1999)}]{free_fermion_Peschel_1999}%
  \BibitemOpen
  \bibfield  {author} {\bibinfo {author} {\bibfnamefont {I.}~\bibnamefont
  {Peschel}}\ and\ \bibinfo {author} {\bibfnamefont {M.-C.}\ \bibnamefont
  {Chung}},\ }\bibfield  {title} {\bibinfo {title} {Density matrices for a
  chain of oscillators},\ }\href {https://doi.org/10.1088/0305-4470/32/48/305}
  {\bibfield  {journal} {\bibinfo  {journal} {Journal of Physics A:
  Mathematical and General}\ }\textbf {\bibinfo {volume} {32}},\ \bibinfo
  {pages} {8419} (\bibinfo {year} {1999})}\BibitemShut {NoStop}%
\bibitem [{\citenamefont {Chung}\ and\ \citenamefont
  {Peschel}(2001)}]{Free_fermion_Peschel_2001}%
  \BibitemOpen
  \bibfield  {author} {\bibinfo {author} {\bibfnamefont {M.-C.}\ \bibnamefont
  {Chung}}\ and\ \bibinfo {author} {\bibfnamefont {I.}~\bibnamefont
  {Peschel}},\ }\bibfield  {title} {\bibinfo {title} {Density-matrix spectra of
  solvable fermionic systems},\ }\href
  {https://doi.org/10.1103/PhysRevB.64.064412} {\bibfield  {journal} {\bibinfo
  {journal} {Phys. Rev. B}\ }\textbf {\bibinfo {volume} {64}},\ \bibinfo
  {pages} {064412} (\bibinfo {year} {2001})}\BibitemShut {NoStop}%
\bibitem [{\citenamefont {Cheong}\ and\ \citenamefont
  {Henley}(2002)}]{Free_fermion_Cheng_2002}%
  \BibitemOpen
  \bibfield  {author} {\bibinfo {author} {\bibfnamefont {S.-A.}\ \bibnamefont
  {Cheong}}\ and\ \bibinfo {author} {\bibfnamefont {C.~L.}\ \bibnamefont
  {Henley}},\ }\href {https://doi.org/10.48550/ARXIV.COND-MAT/0206196}
  {\bibinfo {title} {Many-body density matrices for free fermions}} (\bibinfo
  {year} {2002})\BibitemShut {NoStop}%
\bibitem [{\citenamefont {Vidal}\ \emph {et~al.}(2003)\citenamefont {Vidal},
  \citenamefont {Latorre}, \citenamefont {Rico},\ and\ \citenamefont
  {Kitaev}}]{Free_fermion_Vidal_2003}%
  \BibitemOpen
  \bibfield  {author} {\bibinfo {author} {\bibfnamefont {G.}~\bibnamefont
  {Vidal}}, \bibinfo {author} {\bibfnamefont {J.~I.}\ \bibnamefont {Latorre}},
  \bibinfo {author} {\bibfnamefont {E.}~\bibnamefont {Rico}},\ and\ \bibinfo
  {author} {\bibfnamefont {A.}~\bibnamefont {Kitaev}},\ }\bibfield  {title}
  {\bibinfo {title} {Entanglement in quantum critical phenomena},\ }\bibfield
  {journal} {\bibinfo  {journal} {Physical Review Letters}\ }\textbf {\bibinfo
  {volume} {90}},\ \href {https://doi.org/10.1103/physrevlett.90.227902}
  {10.1103/physrevlett.90.227902} (\bibinfo {year} {2003})\BibitemShut
  {NoStop}%
\bibitem [{\citenamefont {Peschel}(2004)}]{free_fermion_Peschel_2004}%
  \BibitemOpen
  \bibfield  {author} {\bibinfo {author} {\bibfnamefont {I.}~\bibnamefont
  {Peschel}},\ }\bibfield  {title} {\bibinfo {title} {On the reduced density
  matrix for a chain of free electrons},\ }\href
  {https://doi.org/10.1088/1742-5468/2004/06/p06004} {\bibfield  {journal}
  {\bibinfo  {journal} {Journal of Statistical Mechanics: Theory and
  Experiment}\ }\textbf {\bibinfo {volume} {2004}},\ \bibinfo {pages} {P06004}
  (\bibinfo {year} {2004})}\BibitemShut {NoStop}%
\bibitem [{\citenamefont {Cheong}\ and\ \citenamefont
  {Henley}(2004)}]{Free_fermion_Cheong_2004}%
  \BibitemOpen
  \bibfield  {author} {\bibinfo {author} {\bibfnamefont {S.-A.}\ \bibnamefont
  {Cheong}}\ and\ \bibinfo {author} {\bibfnamefont {C.~L.}\ \bibnamefont
  {Henley}},\ }\bibfield  {title} {\bibinfo {title} {Many-body density matrices
  for free fermions},\ }\href {https://doi.org/10.1103/PhysRevB.69.075111}
  {\bibfield  {journal} {\bibinfo  {journal} {Phys. Rev. B}\ }\textbf {\bibinfo
  {volume} {69}},\ \bibinfo {pages} {075111} (\bibinfo {year}
  {2004})}\BibitemShut {NoStop}%
\bibitem [{\citenamefont {Latorre}\ and\ \citenamefont
  {Riera}(2009)}]{free_fermion_Latorre_2009}%
  \BibitemOpen
  \bibfield  {author} {\bibinfo {author} {\bibfnamefont {J.~I.}\ \bibnamefont
  {Latorre}}\ and\ \bibinfo {author} {\bibfnamefont {A.}~\bibnamefont
  {Riera}},\ }\bibfield  {title} {\bibinfo {title} {A short review on
  entanglement in quantum spin systems},\ }\href
  {https://doi.org/10.1088/1751-8113/42/50/504002} {\bibfield  {journal}
  {\bibinfo  {journal} {Journal of Physics A: Mathematical and Theoretical}\
  }\textbf {\bibinfo {volume} {42}},\ \bibinfo {pages} {504002} (\bibinfo
  {year} {2009})}\BibitemShut {NoStop}%
\bibitem [{\citenamefont {Peschel}\ and\ \citenamefont
  {Eisler}(2009)}]{Free_fermion_Peschel_2009}%
  \BibitemOpen
  \bibfield  {author} {\bibinfo {author} {\bibfnamefont {I.}~\bibnamefont
  {Peschel}}\ and\ \bibinfo {author} {\bibfnamefont {V.}~\bibnamefont
  {Eisler}},\ }\bibfield  {title} {\bibinfo {title} {Reduced density matrices
  and entanglement entropy in free lattice models},\ }\href
  {https://doi.org/10.1088/1751-8113/42/50/504003} {\bibfield  {journal}
  {\bibinfo  {journal} {Journal of Physics A: Mathematical and Theoretical}\
  }\textbf {\bibinfo {volume} {42}},\ \bibinfo {pages} {504003} (\bibinfo
  {year} {2009})}\BibitemShut {NoStop}%
\bibitem [{\citenamefont {Fidkowski}(2010)}]{Free_fermion_Fidkowski_2010}%
  \BibitemOpen
  \bibfield  {author} {\bibinfo {author} {\bibfnamefont {L.}~\bibnamefont
  {Fidkowski}},\ }\bibfield  {title} {\bibinfo {title} {Entanglement spectrum
  of topological insulators and superconductors},\ }\href
  {https://doi.org/10.1103/PhysRevLett.104.130502} {\bibfield  {journal}
  {\bibinfo  {journal} {Phys. Rev. Lett.}\ }\textbf {\bibinfo {volume} {104}},\
  \bibinfo {pages} {130502} (\bibinfo {year} {2010})}\BibitemShut {NoStop}%
\bibitem [{\citenamefont {Wannier}(1937)}]{Wannier_Wannier1937}%
  \BibitemOpen
  \bibfield  {author} {\bibinfo {author} {\bibfnamefont {G.~H.}\ \bibnamefont
  {Wannier}},\ }\bibfield  {title} {\bibinfo {title} {The structure of
  electronic excitation levels in insulating crystals},\ }\href
  {https://doi.org/10.1103/PhysRev.52.191} {\bibfield  {journal} {\bibinfo
  {journal} {Phys. Rev.}\ }\textbf {\bibinfo {volume} {52}},\ \bibinfo {pages}
  {191} (\bibinfo {year} {1937})}\BibitemShut {NoStop}%
\bibitem [{\citenamefont {Kivelson}(1982)}]{Wanier_PhysRevB.26.4269}%
  \BibitemOpen
  \bibfield  {author} {\bibinfo {author} {\bibfnamefont {S.}~\bibnamefont
  {Kivelson}},\ }\bibfield  {title} {\bibinfo {title} {Wannier functions in
  one-dimensional disordered systems: Application to fractionally charged
  solitons},\ }\href {https://doi.org/10.1103/PhysRevB.26.4269} {\bibfield
  {journal} {\bibinfo  {journal} {Phys. Rev. B}\ }\textbf {\bibinfo {volume}
  {26}},\ \bibinfo {pages} {4269} (\bibinfo {year} {1982})}\BibitemShut
  {NoStop}%
\bibitem [{\citenamefont {Marzari}\ and\ \citenamefont
  {Vanderbilt}(1997)}]{Wannier_PhysRevB.56.12847}%
  \BibitemOpen
  \bibfield  {author} {\bibinfo {author} {\bibfnamefont {N.}~\bibnamefont
  {Marzari}}\ and\ \bibinfo {author} {\bibfnamefont {D.}~\bibnamefont
  {Vanderbilt}},\ }\bibfield  {title} {\bibinfo {title} {Maximally localized
  generalized wannier functions for composite energy bands},\ }\href
  {https://doi.org/10.1103/PhysRevB.56.12847} {\bibfield  {journal} {\bibinfo
  {journal} {Phys. Rev. B}\ }\textbf {\bibinfo {volume} {56}},\ \bibinfo
  {pages} {12847} (\bibinfo {year} {1997})}\BibitemShut {NoStop}%
\bibitem [{\citenamefont {Marzari}\ \emph {et~al.}(2012)\citenamefont
  {Marzari}, \citenamefont {Mostofi}, \citenamefont {Yates}, \citenamefont
  {Souza},\ and\ \citenamefont {Vanderbilt}}]{Wannier_RevModPhys.84.1419}%
  \BibitemOpen
  \bibfield  {author} {\bibinfo {author} {\bibfnamefont {N.}~\bibnamefont
  {Marzari}}, \bibinfo {author} {\bibfnamefont {A.~A.}\ \bibnamefont
  {Mostofi}}, \bibinfo {author} {\bibfnamefont {J.~R.}\ \bibnamefont {Yates}},
  \bibinfo {author} {\bibfnamefont {I.}~\bibnamefont {Souza}},\ and\ \bibinfo
  {author} {\bibfnamefont {D.}~\bibnamefont {Vanderbilt}},\ }\bibfield  {title}
  {\bibinfo {title} {Maximally localized wannier functions: Theory and
  applications},\ }\href {https://doi.org/10.1103/RevModPhys.84.1419}
  {\bibfield  {journal} {\bibinfo  {journal} {Rev. Mod. Phys.}\ }\textbf
  {\bibinfo {volume} {84}},\ \bibinfo {pages} {1419} (\bibinfo {year}
  {2012})}\BibitemShut {NoStop}%
\bibitem [{\citenamefont {Bradlyn}\ and\ \citenamefont
  {Iraola}(2022)}]{Wannier_Bradlyn_2022}%
  \BibitemOpen
  \bibfield  {author} {\bibinfo {author} {\bibfnamefont {B.}~\bibnamefont
  {Bradlyn}}\ and\ \bibinfo {author} {\bibfnamefont {M.}~\bibnamefont
  {Iraola}},\ }\bibfield  {title} {\bibinfo {title} {Lecture notes on berry
  phases and topology},\ }\bibfield  {journal} {\bibinfo  {journal} {{SciPost}
  Physics Lecture Notes}\ }\href
  {https://doi.org/10.21468/scipostphyslectnotes.51}
  {10.21468/scipostphyslectnotes.51} (\bibinfo {year} {2022})\BibitemShut
  {NoStop}%
\bibitem [{\citenamefont {\ifmmode \mbox{\c{S}}\else
  \c{S}\fi{}ahino\ifmmode~\breve{g}\else \u{g}\fi{}lu}\ \emph
  {et~al.}(2018)\citenamefont {\ifmmode \mbox{\c{S}}\else
  \c{S}\fi{}ahino\ifmmode~\breve{g}\else \u{g}\fi{}lu}, \citenamefont {Shukla},
  \citenamefont {Bi},\ and\ \citenamefont {Chen}}]{MP_rep}%
  \BibitemOpen
  \bibfield  {author} {\bibinfo {author} {\bibfnamefont {M.~B.}\ \bibnamefont
  {\ifmmode \mbox{\c{S}}\else \c{S}\fi{}ahino\ifmmode~\breve{g}\else
  \u{g}\fi{}lu}}, \bibinfo {author} {\bibfnamefont {S.~K.}\ \bibnamefont
  {Shukla}}, \bibinfo {author} {\bibfnamefont {F.}~\bibnamefont {Bi}},\ and\
  \bibinfo {author} {\bibfnamefont {X.}~\bibnamefont {Chen}},\ }\bibfield
  {title} {\bibinfo {title} {Matrix product representation of locality
  preserving unitaries},\ }\href {https://doi.org/10.1103/PhysRevB.98.245122}
  {\bibfield  {journal} {\bibinfo  {journal} {Phys. Rev. B}\ }\textbf {\bibinfo
  {volume} {98}},\ \bibinfo {pages} {245122} (\bibinfo {year}
  {2018})}\BibitemShut {NoStop}%
\end{thebibliography}%

\appendix
\section{Entanglement Bound for frozen modes vs. courier modes \label{app:bound}}
In this appendix, we relate the local threshold $\epsilon_k$ at the $k$-th ZER step to the global entanglement quantifying the approximate factorization of $|\Psi^{[k]}_{courier} \rangle $ into a product state $  |\Psi^{[k+1]}_{courier} \rangle \otimes |\Psi^{[k+1]}_{frozen} \rangle $ between the new courier and frozen modes. 
Here, the frozen modes include both the filled and empty modes. For simplicity, in the following we drop the index corresponding to the RG time, and simply refer to $|\Psi^{[k]}_{courier} \rangle $ as the input state, and the courier, filled, and empty modes correspond to those identified within the current Hilbert space. 

Let us consider the empty modes first. We are interested in the entanglement entropy of the input state with respect to the biparition of the system into the set of empty modes and its complement. 
In our free-fermion context, this entanglement entropy can be obtained as follows: let $\psi_{e}$ be a (rectangular) matrix with the columns being an orthonormal basis for the empty modes. As discussed in the main text, the entanglement entropy is then given by the restricted correlation matrix $C_e^T =  \psi_{e}^\dagger C^T \psi_{e}$  \cite{Free_fermion_Cheng_2002,Free_fermion_Cheong_2004,Free_fermion_Fidkowski_2010,free_fermion_Latorre_2009,Free_fermion_Lieb_1961,free_fermion_Peschel_1999,Free_fermion_Peschel_2001,Free_fermion_Peschel_2003,free_fermion_Peschel_2004,Free_fermion_Peschel_2009,Free_fermion_Vidal_2003}, where $C$ is the correlation matrix of the input state.
Our strategy is to relate the trace of the restricted correlation matrix $C|_\text{empty}$ to the local distillation threshold $\epsilon$. To this end, we consider another (rectangular) matrix $\phi_e$ formed by aggregating the column vectors $\ket{\phi_{e}}_{i} $ distilled from all the subregions, i.e., the transformed fermion creation operators $\hat f^\dagger_i$
\begin{equation}
(\hat f^{\dagger}_1,\hat f^{\dagger}_2,\cdots,\hat f^{\dagger}_{z_e L}) = (\hat c^{\dagger}_1,\hat c^{\dagger}_2,\cdots,\hat c^{\dagger}_n) \phi_e
\end{equation}
are the empty modes localized to the individual distilling regions in space. 
Note that $z_e$ is the average number of empty modes we distill from each local subregion, and $L$ is the system size at the current RG scale.
Importantly, the overlapping of the distilling regions implies the column vectors in $\phi_e$ may not be orthonormal. We may obtain an orthonormal basis for the column span of $\phi_e$ by performing a singular value decomposition:
\begin{equation}
\phi_e = UDV^{\dagger},
\end{equation}
and we may restrict the unitary matrix $U$ to the columns corresponding to the non-zero singular values in $D$.
For simplicity, we assume here that $D$ is full-rank, i.e., the number of non-zero singular values equals to $z_e L$. Nevertheless, The case of a rank-deficit $\phi_e$ can be handled similarly by the method below.
The span of the columns of this restricted matrix gives the same space as that of $\phi_e$, and so we may identify the restricted version of $U$ as the $\psi_e$, using which we define independent fermion operators $\hat f'_i$:
\begin{equation}
    \begin{split}
        (\hat f'^{\dagger}_1,\hat f'^{\dagger}_2,\cdots,\hat f'^{\dagger}_{z_e L}) 
        = & (c^{\dagger}_1,c^{\dagger}_2,\cdots,c^{\dagger}_n)\psi_{\text{e}}\\
        =& (\hat f^{\dagger}_1,\hat f^{\dagger}_2,\cdots,\hat f^{\dagger}_{z_e L}) V D^{-1}\\
        \implies \hat f^{\dagger}_i =& \sum_{l}\hat{f}'^{\dagger}_{l}\lambda_l V^{*}_{il},
    \end{split}
\end{equation}
where $\lambda_l$ are the non-zero diagonal elements of $D$.
Now we consider the trace of the restricted correlation matrix.
\begin{equation}
    \begin{split}
        \sum_i \langle \hat f^{\dagger}_i \hat f_i \rangle &= \sum_{i}\sum_{lm}\langle\hat{f}'^{\dagger}_{l}\lambda_l V^{\dagger}_{li} V_{im}\lambda_m \hat{f}'_m\rangle\\
        &= \sum_{lm}\langle\hat{f}'^{\dagger}_{l}\lambda_l \delta_{lm}\lambda_m \hat{f}'_m\rangle\\
        &= \sum_{l}\lambda_l^{2} \langle\hat{f}'^{\dagger}_{l} \hat{f}'_l\rangle\\
        \implies \epsilon z_e L &\geq \sum_l \lambda_l^{2} \langle \hat{f}'^{\dagger}_l \hat{f}'_l \rangle  \geq \min(\lambda_l^{2}) \sum_l \langle \hat{f}'^{\dagger}_l \hat{f}'_l \rangle \\
        \implies \frac{\epsilon }{\min(\lambda_l^{2})} z_e L &\geq {\rm Tr}(C_e).
    \end{split}
\end{equation}

Now, we use the bound on ${\rm Tr}(C_e)$ to bound the entanglement entropy $S_e$. Recall
\begin{equation}
    S_{e}  
= \sum_l S(\xi_l) 
= -\left(\sum_{l} \xi_{l} \ln(\xi_{l}) + (1 - \xi_{l}) \ln(1 - \xi_{l}) \right),
\end{equation}
where $\xi_n$ denotes the eigenvalues of the restricted correlation matrix $C_e$. Since $S$ is concave, we have
\begin{equation}\begin{split}\label{eq:}
\frac{1}{z_e L }\sum_l S(\xi_l)  \leq S \left( \frac{\sum_{l} \xi_l}{z_e L} \right) = S \left( \frac{{\rm Tr}(C_e)}{z_e L} \right).
\end{split}\end{equation}
As $S$ attains maximum at $x=\frac{1}{2}$ and increases monotonically in the range $(0,\frac{1}{2}]$, if $\frac{\epsilon }{\min(\lambda_l^{2})} \leq 1/2$, we have
\begin{equation}
S \left( \frac{{\rm Tr}(C_e)}{z_e L} \right) \leq S \left( \frac{\epsilon }{\min(\lambda_l^{2})} \right).
\end{equation}
This establishes the bound
\begin{equation}
S_{e}  
\leq z_e S \left( \frac{\epsilon }{\min(\lambda_l^{2})} \right)  L .
\end{equation}

With a similar argument, let us consider the filled modes. The basis rotation matrices $\psi_f$ and $\phi_f$ are defined similarly. However, instead of using the correlation matrix directly we alternatively consider $1-C_{f}$, i.e., we consider $\sum_i \langle \hat f_i \hat f^{\dagger}_i \rangle$ instead. The previous arguments apply in the same way, and we conclude
\begin{equation}
S_{f}  
\leq z_f S \left( \frac{\epsilon }{\min(\mu_l^{2})} \right)  L,
\end{equation}
where $\mu_l$ denotes the singular values arising from the orthonormalization of $\phi_f$.

Combining the two bounds above, we may bound the entanglement entropy with respect to the separation of courier vs.\ frozen mode. Since the input state is pure, the strong subaddivity of entanglement entropy implies
\begin{equation}\begin{split}\label{eq:}
S_\text{courier} \leq&  S_{e} + S_{f}\\
\leq & \left( z_e S \left( \frac{\epsilon }{\min(\lambda_l^{2})}\right) + z_f S \left( \frac{\epsilon }{\min(\mu_l^{2})}\right) \right) L.
\end{split}\end{equation}
This establishes the bound claimed in the main text.

\begin{figure}[h]
\begin{center}
{\includegraphics[width=0.48 \textwidth]{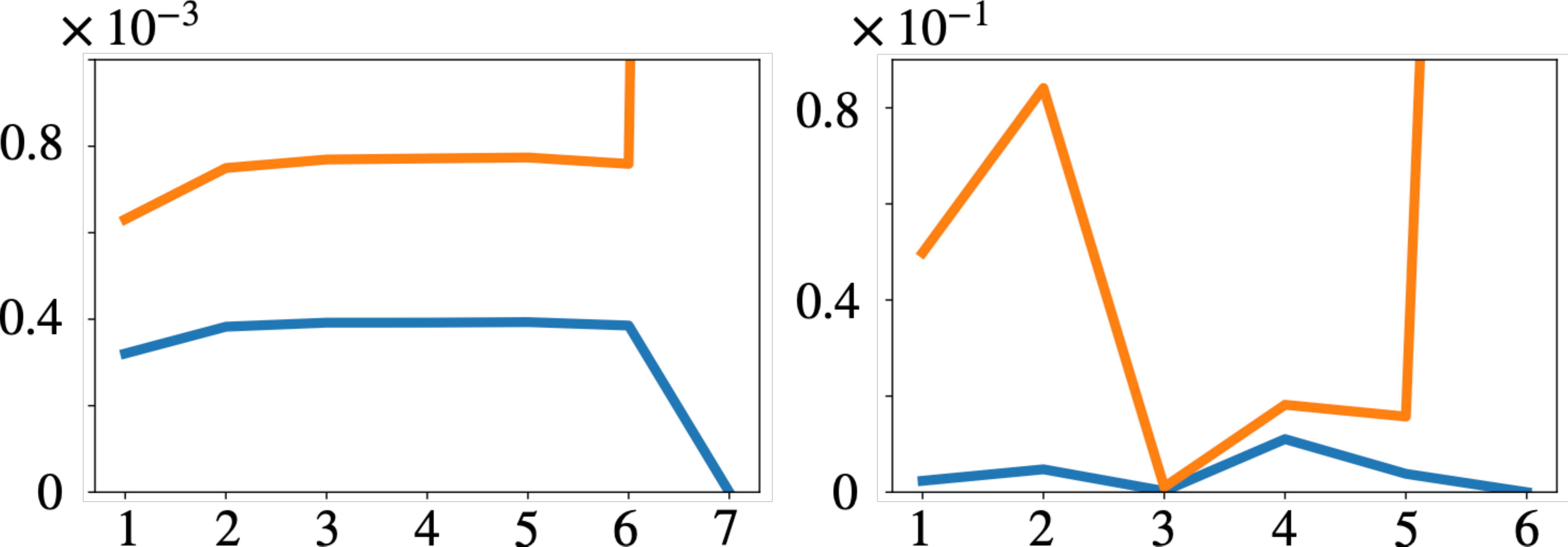}} 
\caption{Plot of entanglement entropy density $S_\text{courier}/L$ (blue) vs.\ the bound in Eq.\ \eqref{eq:S_bound} (orange) against the ZER steps. 
\label{fig:EE_density}
 }
\end{center}
\end{figure}

For the following numeric, to get a tighter bound, we use $\epsilon_{e} = \max(\langle \hat f^{\dagger}_i \hat f_i \rangle)$ for $S_e$ and $\epsilon_{f} = \max(\langle \hat f_i \hat f^{\dagger}_i \rangle)$ for $S_f$, thus 
\begin{equation}\begin{split}\label{eq:S_bound}
S_\text{courier}  \leq & \left( z_e S \left( \frac{\epsilon_e }{\min(\lambda_l^{2})}\right) + z_f S \left( \frac{\epsilon_f }{\min(\mu_l^{2})}\right) \right) L.
\end{split}\end{equation}
The comparison between the entanglement entropy involved in our ZER process and the bound obtained above is shown in Fig.\ \ref{fig:EE_density}.

\section{Projected Position Operator Method for Wannier Function in 1D
\label{app:WF}}
 One of the main difference between ZER and MERA is the implementation of Wannierization in the zipper unitary. With Wannierization we can find a localized basis for different modes and rotate the correlation into this basis to proceed renormalization process. Wannierization in higher dimensions and relation between topological invariances are discussed in detail in \cite{Wanier_PhysRevB.26.4269, Wannier_Bradlyn_2022, Wannier_PhysRevB.56.12847, Wannier_Wannier1937}. In this section we will introduce Wannierization process of 1D problems which is called projected position method. Suppose the Hamiltonian is translational invariant and it commutes with the translational operator. From Bloch's theorem we can obtain the Bloch waves $| \psi_{n \textbf{k}} \rangle $ which are the simultaneous eigenstates of Hamiltonain and translational operator. 
 These Bloch states are periodic in Brillouin zone and extend over the momentum space. From this perspective, a natural question is to ask is it possible to find the  localized description for the Bloch states and this can be achieved by Wannierization process, finding the Wannier function in real space. The simplest way to obtain the Wannier function is to use Fourier transform
 \begin{equation}
     | \textbf{R} n \rangle = \frac{V}{(2 \pi)^{D}} \int_{BZ} d \textbf{k} ~ e^{i \textbf{k} \cdot \textbf{R}} | \psi_{n \textbf{k}} \rangle,
 \end{equation}
 where $| \textbf{R} n \rangle $ is the Wannier function in real space, $\textbf{R}$ is the lattice vector, and $D$ is the spatial dimension. However, there is no unique definition of Wannier function since we can apply unitary transformation to the Bloch states $| \psi_{n \textbf{k}} \rangle$ which does not change the physics \cite{Wannier_RevModPhys.84.1419, Wannier_PhysRevB.56.12847}. Therefore, finding a suitable gauge to transform the Bloch states and resulting the most localized Wannier function is not trivial. However, in 1D the problem can be much more simpler by looking at the projected position operator $PxP$ \cite{Wannier_PhysRevB.56.12847} . The band projector is defined by the $M$ bands of interests labelled
 \begin{equation}
 	P = \sum^{M}_{n} \sum_{k} | \psi_{nk} \rangle \langle \psi_{nk} | = \sum_{n}^{M} \sum_{R} | R m \rangle \langle R m  | . 
 \end{equation}
The Wannier function is then obtained by computing the eigenvectors of the projected position operator $PxP$, where $x$ is the position operator in real space. The corresponding eigenvalues are the "Wannier charge center" . Besides, these Wannier function is maximally localized in real space since they minimize the spread functional. In this perspective, we call such eigenvectors as "Maximally localized Wannier Function(MLWF)". One should be cautious that this method does not work in 2D since the projected position operator in other direction(i.e. $PyP, PzP$) does not commute each other \cite{Wannier_PhysRevB.56.12847}.  

\section{Details of the numerics}
The details on the numerical implementation of ZER in the three models discussed in the main text are summarized in Table \ref{table:1}.

\begin{center}
\begin{table}[h!]
\begin{tabular}{|c|c|c|c|}
 \hline
 Model & SSH &\begin{tabular}[c]{@{}c@{}} Nearest \\ neighbour \end{tabular} & Extended \\ \hline
 \begin{tabular}[c]{@{}l@{}}System size\\ (\# unit cell) \end{tabular} & 729 & 1024 & 1024 \\ \hline
 Threshold ($\epsilon$) &
 $10^{-5} $ &   $10^{-4} $  & * \\ \hline 
 Model parameter & \begin{tabular}[c]{@{}l@{}} $t_1 = -0.4$ \\ $t_2 = -0.6$ \end{tabular} & $t=-1$ & \begin{tabular}[c]{@{}l@{}} $t_1 = -1$ \\ $t_2 = -2$ \end{tabular} \\\hline
 \# core modes  & 0 & 8 & 8\\ \hline
\end{tabular}
\caption{Summary of the details of the numeric. For the SSH model, $t_1$ and $t_2$ are respectively the intra- and inter-unit cell electron hopping amplitudes. For the extended model, $t_1$ and $t_2$ are respectively the nearest and next-nearest neighbor hopping amplitudes. In addition, we adjusted the local distillation threshold in the ZER process to minimize the number of core modes left, 
and * denotes $5\times 10^{-4}$,$5\times 10^{-4}$,$5\times 10^{-4}$, $10^{-3}$, $10^{-3}$, and $10^{-2}$ from early to late RG times.
}
\label{table:1}
\end{table}
\end{center}

\end{document}